\documentclass[%
 reprint,
superscriptaddress,
showpacs,preprintnumbers,
showkeys,
amsmath,amssymb,
aps,
pra,
]{revtex4-1}

\usepackage{graphicx}% Include figure files
\usepackage{dcolumn}% Align table columns on decimal point
\usepackage{bm}% bold math
\usepackage{braket}
\begin{document}

\preprint{APS/123-QED}

\title{Generation of energy-entangled W states via parametric fluorescence\\in integrated devices}

\author{M. Menotti}
 \email{matteo.menotti01@ateneopv.it}
 \affiliation{Department of Physics, University of Pavia, Via Bassi 6, I-27100 Pavia, Italy}
\author{L. Maccone}
 \affiliation{Department of Physics, University of Pavia, Via Bassi 6, I-27100 Pavia, Italy}
 \affiliation{INFN Sezione di Pavia, University of Pavia, Via Bassi 6, I-27100 Pavia, Italy}
\author{J. E. Sipe}
 \affiliation{Department of Physics, University of Toronto, 60 St. George Street, Toronto,\\ Ontario M5S 1A7, Canada}
\author{M. Liscidini}
 \affiliation{Department of Physics, University of Pavia, Via Bassi 6, I-27100 Pavia, Italy}

\date{\today}

\begin{abstract}
Tripartite entangled states, such as GHZ and W states, are typically generated by manipulating two pairs of polarization-entangled photons in bulk optics.
Here we propose a scheme to generate W states that are entangled in the energy degree of freedom in an integrated optical circuit.
Our approach employs photon pairs generated by spontaneous four-wave mixing (SFWM) in a micro-ring resonator.
We also present a feasible procedure for demonstrating the generation of such a state, and we compare polarization-entangled and energy-entangled schemes for the preparation of W states.
\end{abstract}

%Use showkeys class option if keyword
\maketitle

Entanglement is one of the most characteristic and interesting features of quantum mechanics, and it has been extensively studied both from the perspective of fundamental physics and for applications in quantum information processing.
To this end, tests and protocols involving entangled states have been devised, which have led to an improved understanding of their properties, and to major achievements in quantum cryptography and teleportation.
Originally entangled states of bipartite systems were studied, and today they are well understood \cite{Nielsen,Horodecki,Plenio}.
Entangled states of composite systems involving more than two subsystems, or \emph{multipartite entangled states}, are a more recent focus of attention \cite{GHZ_Paper,Zeilinger,Horodecki,Schwaiger}.
In general the correlations they exhibit cannot be considered a trivial generalization of bipartite entanglement.

Tripartite entangled states are divided into two inequivalent classes \cite{Dur,GHZ_Book,GHZ_Paper,Horodecki,Schwaiger}, represented by Greenberger-Horne-Zeilinger (GHZ) states and W states, which cannot be converted into each other using stochastic local operations and classical communication.
Both types of states display interesting features. For example, the GHZ states have been involved in the demonstration of Bell's inequalities violation \cite{Pan1}, as well as in teleportation protocols \cite{Pan2} and in super dense coding \cite{Hao}.
The W states, while perhaps less studied, have been shown to be promising candidates for the implementation of a range of quantum protocols \cite{Shi,Yamamoto,Eibl,Kiesel}.
Moreover, they are more robust with respect to losses than GHZ states \cite{Kiesel}.

The W state \cite{Dur} has the form 
\begin{equation}\label{W paradigm}
	\ket{W}=\frac{1}{\sqrt{3}}\big(\ket{001}+\ket{010}+\ket{100}\big),
\end{equation}
where $\ket{0}$ and $\ket{1}$ indicate orthogonal states. Typically they refer to either ``vacuum'' and ``occupation'', or ``ground state'' and ``excited state''.
In the former instance, (\ref{W paradigm}) would involve only one particle.
However, many protocols involving W states require the presence of three actual particles, such as ions or photons. 
This is a necessary condition, for instance, in tests of non-locality and to exploit the robustness of W states in quantum communications.

Photons in a tripartite system have been entangled in their polarization degree of freedom, so that a three-photon W state might take the form
\begin{equation}\label{W polarization}
	\ket{W}=\frac{1}{\sqrt{3}}\big(\ket{HHV}+\ket{HVH}+\ket{VHH}\big),
\end{equation}
where $H$ and $V$ indicate horizontal and vertical polarization, respectively.
The experimental preparation of such states usually requires the simultaneous generation of two photon pairs by means of type-II spontaneous parametric down-conversion (SPDC) in a bulk nonlinear crystal \cite{Yamamoto,Eibl}. The photons then propagate in free space or in optical fibers, and one can perform operations on them by using optical elements such as beam splitters, polarization beam splitters, and wave-plates.

In this communication we propose a scheme to generate W states that relies on energy-entanglement rather than polarization entanglement.
Our scheme can be implemented in an integrated optics platform, and thus has the advantages of scalability and efficiency.

\begin{figure}
\includegraphics[scale=0.16,keepaspectratio]{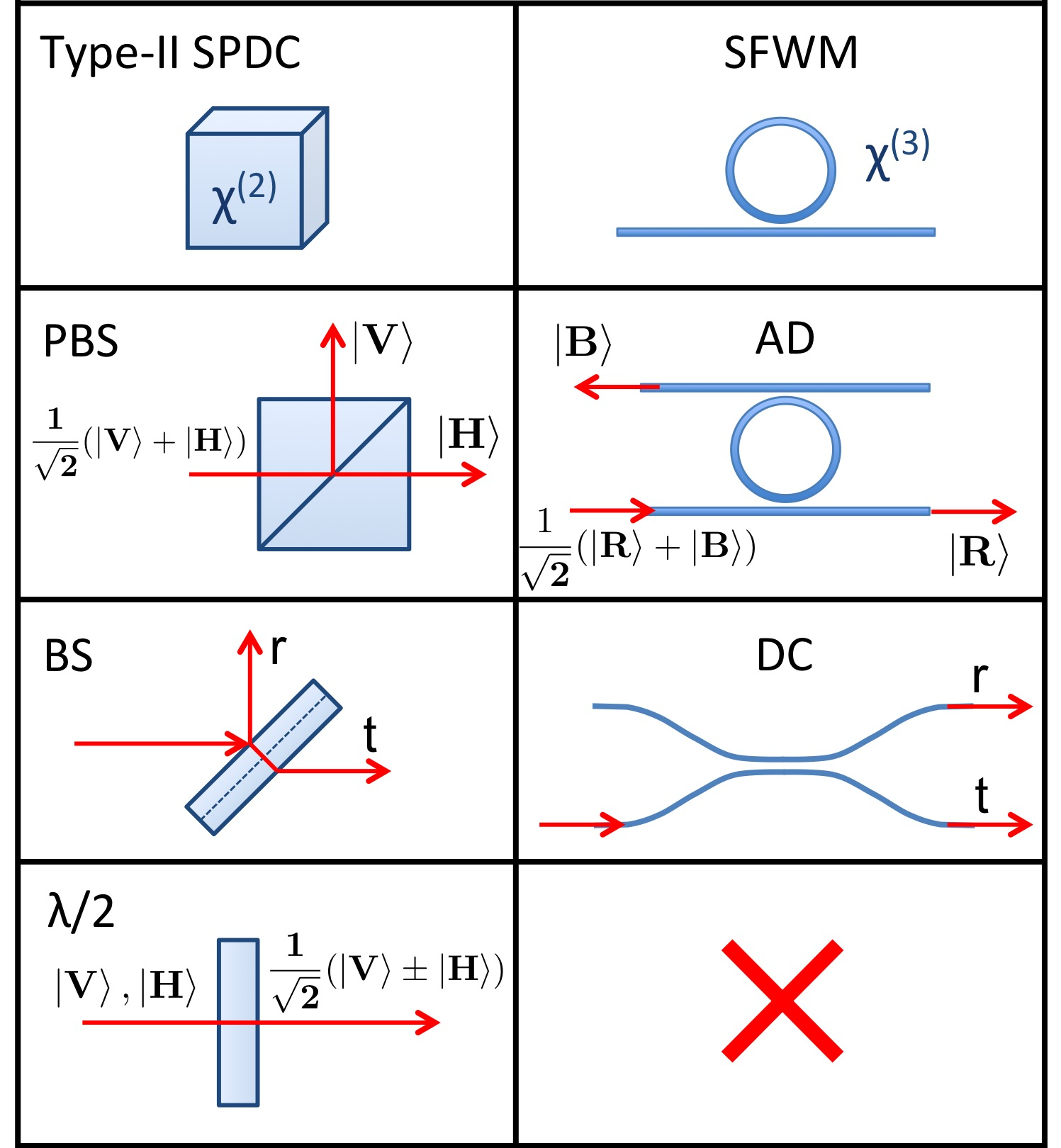}
\caption{\label{Analogies}Analogies between optical elements employed in bulk optics for schemes involving polarization-entangled states (on the left) and the corresponding integrated optical elements for the scheme introduced here involving energy-entangled states (on the right). Note the absence of an analogue for the $\lambda/2$ waveplate.}
\end{figure}

The scheme presented here prepares a W state of the form
\begin{equation}
\frac{1}{\sqrt{3}}(\ket{BBR}+\ket{BRB}+\ket{RBB})
\end{equation}
or
\begin{equation}
\frac{1}{\sqrt{3}}(\ket{RRB}+\ket{RBR}+\ket{BRR}),
\end{equation}
where $\ket{B}$ and $\ket{R}$ are photons that are blue and red detuned with respect to the pump.
Before presenting the specific structure we have in mind, we want to compare the bulk optical elements used to manipulate polarization-entangled states with the integrated optical elements used to manipulate energy-entangled states. This allows us to establish a correspondence, when possible, which is particularly useful in translating a polarization- to an energy-entanglement scheme for the generation of W states.
This correspondence is shown in Fig. \ref{Analogies}.

The sources commonly considered for the preparation of polarization-entangled photon pairs in bulk devices are nonlinear crystals such as $\beta$ Barium Borate (BBO), in which type-II SPDC is used to generate the photons; instead a source for the preparation of energy-entangled photon pairs in an integrated structure is a silicon micro-ring resonator, in which spontaneous four-wave mixing (SFWM) is used to generate the photons \cite{Bajoni2}.
Depending on their polarization, photons generated in a bulk crystal can be spatially separated using a polarization beam splitter (PBS); instead, photons generated in a micro-ring resonator can be spatially separated depending on their energy using a tunable add-drop filter \cite{Heebner}.
Finally, the beam splitters commonly employed for photons generated in bulk optics can be replaced by directional couplers for photons generated in integrated structures \cite{Heebner}.

However, not all of the elements in polarization-entanglement optics find a straightforward analog in energy-entanglement integrated optics. For example, in the work by Bouwmeester et al. \cite{GHZ_Paper} on the preparation of polarization-entangled GHZ states, a $\lambda/2$ plate is used to rotate the polarization.
This element cannot be replaced with a linear component in an integrated optical scheme such as that proposed here, for it would require a change in the photon energy.
Although this feature seems detrimental to the design of a source of energy-entangled photons, performing such operations is not strictly required to produce a W state. Moreover, we show in the following that this is not a necessary condition even to obtain a full reconstruction of the density matrix of the state.

\begin{figure*}
\includegraphics[width=\textwidth,height=\textheight,keepaspectratio]{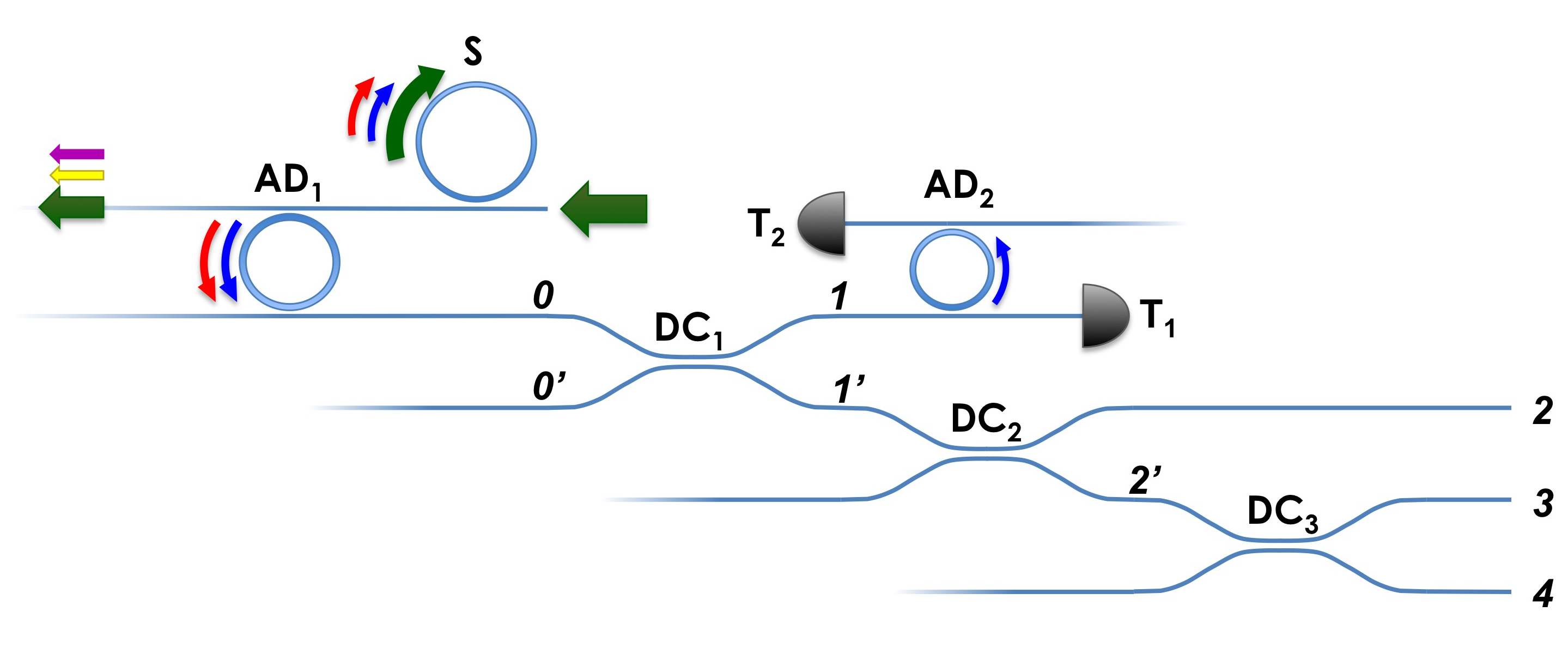}
\caption{\label{circuit}Scheme for generating energy-entangled W states. $S$) Micro-ring resonator for the production of photons pairs by SFWM. $AD_1$) Add-drop filter extracting only the red and blue photons. $AD_2$) Add-drop filter routing the red and blue photons to different target detectors. The $DC_i$s are directional couplers, characterized by real-valued transmission and reflection coefficients $t_i$ and $r_i$, respectively.}
\end{figure*}

We consider the scheme sketched in Fig. \ref{circuit}. It is composed of a silicon micro-ring resonator acting as the source ($S$) for the generation of photon pairs, two add-drop filters ($AD_1$ and $AD_2$) for the extraction of specific spectral components, and three cascaded directional couplers (referred to as $DC_1$, $DC_2$, and $DC_3$).

An intense pump pulse is injected into the silicon micro-ring resonator S, with the center frequency of the pulse chosen to equal that of one of the resonances of the micro-ring.
The small volume of the cavity, together with the build-up of the field due to constructive interference in the ring, dramatically enhances the nonlinear response of the resonator \cite{Helt2}.
Thus we can safely assume that the pair generation takes place in the ring resonator alone.
If the generation rate is low and the undepleted pump approximation is assumed, the state produced by SFWM can be expressed as
\begin{equation}\label{input state}
	\ket{\psi}=\Big(1-\frac{1}{2}\beta^2\Big)\ket{vac}+\beta C_{II}^\dagger\ket{vac}+\frac{1}{2}[\beta C_{II}^\dagger]^2\ket{vac}+\dots,
\end{equation}
where $C_{II}^\dagger$ is the creation operator for a photon pair, and $|\beta|^2$ is its generation probability.
We have truncated the expansion (\ref{input state}) at the second order, which corresponds to the creation of two photon pairs, and we have assumed negligible time-ordering corrections \cite{Sipe}.
In order to generate a W state we consider only the creation of two photon pairs; the probability of this event is $|\beta|^4$.
In particular, here we consider SFWM in a ring resonator when one of the generated photons in each pair is centred at a resonant frequency below that of the pump, and one at a resonant frequency above.
While pairs of photons can be generated at many different resonant frequencies, as long as the energy is conserved and the phase matching condition is satisfied, we focus on energy-entangled W states in which photons are generated via two particular resonances, that we will assume as our $\ket{R}$ and $\ket{B}$ photons; we imagine discarding every other photon that could enter the rest of the circuit.

To do this, one could simply place notch filters in front of each detector, but in the spirit of integration we consider using another ring resonator in the configuration of an add-drop filter ($AD_2$).
The radius of this ring is chosen to guarantee the resonant condition only for red and blue photons; we can obtain this result by choosing a radius that is incommensurable with that of the ring $S$.
Ideally, this solution should also remove the pump photons, but in practice an additional filtering stage may be necessary \cite{Bajoni}.

It has been demonstrated by Helt et al. \cite{Helt} that in SFWM in micro-ring resonators the energy correlation of the photons depends on the pump pulse duration, and it can be tailored to generate a variety of states, ranging from strongly correlated to nearly uncorrelated photon pairs.
Here we focus our attention on the case of a very short pump pulse, since we are not interested in the energy correlations within a single photon pair, but on the entanglement resulting from the manipulation of two identical photon pairs.
Thus we can recast the expression for the state produced by the nonlinear interaction in the micro-ring as
\begin{equation}\label{input state - double pair}
	\ket{\psi}=\beta^2(a_{B,0}^\dagger a_{R,0}^\dagger)^2\ket{vac},
\end{equation}
where $a_{R,0}^\dagger$ and $a_{B,0}^\dagger$ are the creation operators for the red and blue photons, respectively, in channel $0$.

Once this state is identified, we can calculate the output state following the linear propagation of the fields inside the structure.
As suggested by the asymptotic field approach \cite{Liscidini}, the output state is obtained by expressing the input creation operators in (\ref{input state - double pair}) in terms of the creation operators associated with each output channel. The link between the operators is provided by the scattering matrix of the structure.
For instance, considering only the first directional coupler $DC_1$ in Fig. \ref{circuit}, the creation operators for a blue photon in channel $0$ and $0'$, namely $a_{B,0}^\dagger$ and $a_{B,0'}^\dagger$, can be expressed in terms of the creation operators in channel $1$ and $1'$ as
\begin{equation}\label{}
	\begin{bmatrix}
         a_{B,0}^\dagger \\
         a_{B,0'}^\dagger
         \end{bmatrix}
         =\begin{bmatrix}
	t_1 & r_1e^{i\phi_1} \\
	r_1e^{i\phi_1} & t_1
	\end{bmatrix}
	\begin{bmatrix}
	a_{B,1}^\dagger \\
	a_{B,1'}^\dagger
	\end{bmatrix}
\end{equation}
where $r_1$ and $t_1$ are real reflection and transmission coefficients, satisfying $r_1^2+t_1^2=1$, $\phi_1$ is a relative phase shift, and $a_{B,1}^\dagger$ and $a_{B,1'}^\dagger$ are the photon creation operators in the two output channels.

Once the path between the ring and the detector is stabilized with the appropriate relative phase factors, from an asymptotic field approach \cite{Liscidini} we find the output state is
\begin{align}\label{output state}
	\ket{\psi}&=\alpha^2\ket{\Phi}+4\sqrt{3}\beta^2r_1t_1^3r_2t_2^2r_3t_3e^{i(3\phi_1+2\phi_2+\phi_3)}\\
	&\times \Big[a_{R,1}^\dagger\ket{W_{T_1}}+a_{B,1}^\dagger\ket{W_{T_2}}\Big],\nonumber
\end{align}
where
\begin{align}\label{W_T_1}
	\ket{W_{T_1}}&=\frac{1}{\sqrt{3}}\Big[a_{B,2}^\dagger a_{B,3}^\dagger a_{R,4}^\dagger+a_{B,2}^\dagger a_{R,3}^\dagger a_{B,4}^\dagger\\
	&+a_{R,2}^\dagger a_{B,3}^\dagger a_{B,4}^\dagger\Big]\ket{vac}\nonumber
\end{align}
and
\begin{align}\label{W_T_2}
	\ket{W_{T_2}}&=\frac{1}{\sqrt{3}}\Big[a_{R,2}^\dagger a_{R,3}^\dagger a_{B,4}^\dagger+a_{R,2}^\dagger a_{B,3}^\dagger a_{R,4}^\dagger\\
	&+ a_{B,2}^\dagger a_{R,3}^\dagger a_{R,4}^\dagger\Big]\ket{vac}\nonumber
\end{align}
are two normalized W states.
In (\ref{output state}), the state vector $\ket{\Phi}$ includes all the terms that do not involve a single photon in each of the four output channels, and $\alpha$ is a complex constant.
So the state obtained in the three channels ($2$,$3$,$4$) is $\ket{W_{T_1}}$ or $\ket{W_{T_2}}$, depending on the energy of the photon in channel $1$.

To discriminate between $\ket{W_{T_1}}$ and $\ket{W_{T2}}$, we introduce an add-drop filter ($AD_2$) that routes red and blue photons to different target detectors $T_1$ and $T_2$.
The radius of the micro-ring resonator in the add-drop filter configuration is chosen to guarantee a resonance for the blue photons, and high transmission probability for the red photons.
In the present scheme we set the $AD_2$ micro-ring radius to be $1/4$ that of the radius of the $S$ micro-ring, so that the energy of a red photon is found in the middle of the free spectral range of the $AD_2$ ring.
Both detection events occur with probability $|4\sqrt{3}\beta^2r_1t_1^3r_2t_2^2r_3t_3|^2$. The maximum is found for $r_1=\frac{1}{2}$, $r_2=\frac{1}{\sqrt{3}}$, and $r_3=\frac{1}{\sqrt{2}}$, corresponding to a W state generation probability of $|\frac{3}{16}\beta^2|^2$.

To  demonstrate generation of a W state, we could perform a series of measurements in the output channels $2$, $3$, and $4$.
Let us focus, for instance, on the demonstration of the $W_{T_1}$ state in (\ref{W_T_1}).
We consider the post-selection of the event characterized by the simultaneous detection of a red photon in channel $1$, and one photon in each channel $2$, $3$, and $4$.

\begin{figure}
\includegraphics[scale=0.1, keepaspectratio]{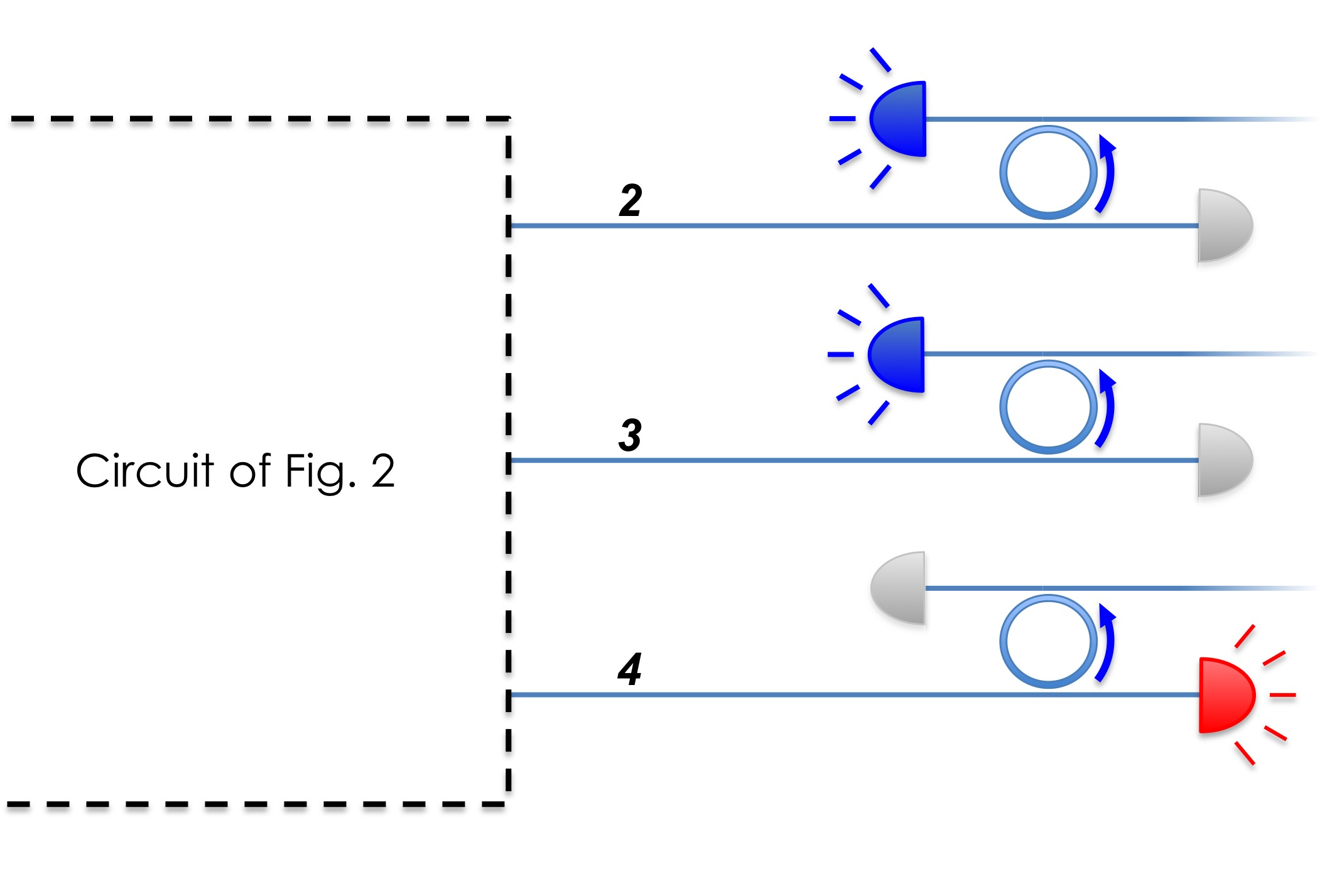}
\caption{\label{detection}Schematic representation of the detection setup. Here, as an example, we sketch a threefold coincidence measurement revealing the $\ket{BBR}$ state in the output channels. All these measurements are subject to the simultaneous detection of a red photon by $T_1$ (see Fig. \ref{circuit}). Note that not all of the detectors represented are actually required: the density matrix of the state can be fully reconstructed using four detectors.}
\end{figure}

Since we need to evaluate the energy of the photons in the output channels, we could employ three add-drop filters identical to $AD_2$, one for each channel, as sketched in Fig. \ref{detection}. Their role is to route the incoming photons on the basis of their energy to different frequency-independent detectors.

The first step toward proving the preparation of the W (\ref{W_T_1}) state is to ensure that the states
\begin{align}\label{notation}
\ket{BBR}\equiv a_{B,2}^\dagger a_{B,3}^\dagger a_{R,4}^\dagger\ket{vac}\\
\ket{BRB}\equiv a_{B,2}^\dagger a_{R,3}^\dagger a_{B,4}^\dagger\ket{vac}\\
\ket{RBB}\equiv a_{R,2}^\dagger a_{B,3}^\dagger a_{B,4}^\dagger\ket{vac}
\end{align}
occur with the same $\frac{1}{3}$ probability. This can be verified by counting the coincidences in the three detectors in $2$, $3$, and $4$. In the following we will assume that this condition has been verified.

As pointed out by Ac\'{i}n et al. \cite{Acin} and more recently by Eibl et al. \cite{Eibl}, this is not a sufficient condition to confirm the generation of a W state. Indeed, a variety of mixed states lead to the same statistics in the coincidence count, such as the incoherent mixture $\rho_{S}=\frac{1}{3}\{\ket{BBR}\bra{BBR}+\ket{BRB}\bra{BRB}+\ket{RBB}\bra{RBB}\}$ and a mixture of biseparable states $\rho_{B}=\frac{1}{3}\{\rho_2\otimes \rho_{34}+\rho_3\otimes \rho_{24}+\rho_4\otimes \rho_{23}\}$ where, for example, $\rho_2$ corresponds to a blue photon in channel $2$, and $\rho_{34}$ represents a Bell state in channels $3$ and $4$.

To rule out these possibilities we can perform full tomography on the state generated by our structure as described below.
Whenever $\ket{RRB}$, $\ket{RBR}$, and $\ket{BRR}$ occur with equal probability, the density matrix describing a system composed of two blue photons and one red photon distributed in channels $2$, $3$, and $4$ is
\begin{equation}\label{three-photon}
\rho_{234} = 
    \bordermatrix{ & \bra{BBR} & \bra{BRB} & \bra{RBB} \cr
      \ket{BBR} & \frac{1}{3} & a & b \cr
      \ket{BRB} & a^* & \frac{1}{3} & c \cr
      \ket{RBB} & b^* & c^* & \frac{1}{3} } \qquad
\end{equation}
where we have specified the diagonal elements, and $a$, $b$, and $c$ are complex numbers to be determined.
To this end, we could evaluate the state of the photon pair in channels $3$ and $4$, subject to the detection of a blue photon in channel $2$.
Given the detection of a blue photon in channel $2$, the density matrix associated with the photon pair in channels $3$ and $4$ is
\begin{equation}
\rho_{34} =
    \bordermatrix{ & \bra{BR} & \bra{RB} \cr
      \ket{BR} & \frac{1}{2} & \frac{3}{2}a \cr
      \ket{RB} & \frac{3}{2}a^* & \frac{1}{2} }. \qquad
\end{equation}

Of course, the general density matrix for two photons, each either red or blue, and one in each of two channels, is a $4$x$4$ matrix.
In our particular case, since the process creating the photons satisfies energy conservation, we expect only one blue and one red photon.
Hence that larger $4$x$4$ matrix must be of the form
\begin{equation}\label{our_case}
\rho_{pair} = 
   \bordermatrix{ & \bra{BB} & \bra{BR} & \bra{RB} & \bra{RR} \cr
      \ket{BB} & 0 & 0 & 0 & 0 \cr
      \ket{BR} & 0 & \frac{1}{2} & \frac{3}{2}a & 0 \cr
      \ket{RB} & 0 & \frac{3}{2}a^* & \frac{1}{2} & 0 \cr
      \ket{RR} & 0 & 0 & 0 & 0 \cr }. \qquad
\end{equation}
The energy correlation of the two photons responsible for the simple form of $\rho_{pair}$ allows for the determination of $a$ by performing tomography on the photon pairs in channels $3$ and $4$, as was done by Ramelow et al. \cite{Ramelow}; this entails simple interferometric measurements between the two modes.

We could apply the same reasoning to the tomography on channels $2$ and $4$ subject to the detection of a blue photon in channel $3$, which would lead to the determination of the complex coefficient $b$; and to the tomography on channels $2$ and $3$ subject to the detection of a blue photon in channel $4$, which would lead to the determination of the complex coefficient $c$.
This procedure allows us to reconstruct the density matrix (\ref{three-photon}) of the three-photon state, and prove the generation of a $W_{T_{1}}$ state.
The protocol to prove the generation of a $W_{T_{2}}$ state is analogous.

In conclusion, in this communication we have presented a design for a device capable of generating energy-entangled W states, relying on SFWM in a micro-ring resonator and the linear propagation of light in an integrated optical circuit.
All of the elements involved in the integrated structure we propose are feasible with current technology and have already been characterized individually.
We have compared them with the elements used for the generation of polarization-entangled states using bulk optics.
Finally, we have presented a feasible protocol to confirm that a W state is indeed generated when the two pairs of photons are initially produced.
It can be used to test the performance of the source we propose.

\bibliography{text}

\end{document}